\documentstyle[preprint,aps]{revtex} 
\tightenlines 
\begin{document}

\draft
\title{Analytical solution of the relativistic Coulomb problem with a 
hard core interaction for a one-dimensional spinless Salpeter equation}
\author{F. Brau\thanks{Chercheur I.I.S.N.}\thanks{E-mail: 
fabian.brau@umh.ac.be}} 
\address{Universit\'{e} de Mons-Hainaut, Place du Parc 20,
B-7000 Mons, BELGIQUE}
\date{December 1998}

\maketitle

\begin{abstract}

In this paper, we construct an analytical solution of the one-dimensional spinless Salpeter 
equation with a Coulomb potential supplemented by a hard core interaction, which keeps the 
particle in the x positive region.  
\end{abstract}

\section{Introduction}
\label{sec:intro}

A simple relativistic version of the Schr\"{o}dinger equation is the Spinless Salpeter 
Equation~(SSE). For the one-dimensional case we have
\begin{equation}
\label{eq*}
\sqrt{-d^2_x + m^2}\, \Psi(x) = \Bigl(E - V(x)\Bigr)\, \Psi(x),
\end{equation}
where $m$ is the mass of the particle, $V(x)$ is the potential 
interaction, $E$ the eigenenergy of the stationary state $\Psi(x)$, $d^2_x=\frac{d^2}
{dx^2}=-p^2$ and $p$ is the relative momentum of the particle ($\hbar=c=1$). 
$p$ and $x$ are conjugate variables. The differential operator of the 
Schr\"{o}dinger equation is well defined because it is a second derivative. To solve a 
physical problem, we must just solve an ordinary eigenvalue differential equation. The 
situation is more complicated with the SSE because the associated differential operator is a 
nonlocal one. Its 
action cannot be calculated directly from its operator form. Indeed, its action on a function 
$f(x)$ is known only if $f(x)$ is an eigenfunction of the operator $d^2_x$. In this case we 
obtain
\begin{equation}
\label{eq0}
\sqrt{-d^2_x + m^2}\ f(x) = \sqrt{-\alpha + m^2}\ f(x),
\end{equation}
where $\alpha$ is the corresponding eigenvalue of $d^2_x$. That is why we need first to rewrite 
the SSE into a form easier to handle. Since the operator is a nonlocal one, this form 
could be an integral equation. This have been done for the three-dimensional 
case in Refs. \cite{nick84,brau98}. We present the one-dimensional corresponding form in the 
next section. With the method used to obtain this form, it is possible to rewrite the SSE as an 
integro-differential equation (see Ref. \cite{brau98} for the three-dimensional case). 
But the kernel is really complicated and the resulting equation seems to be very difficult 
to treat. We will use, here, another method to obtain the solution of the equation.

To solve the relativistic Coulomb problem we do not solve any differential equation.
We calculate the action of the square-root operator on the functions $x^n e^{-\beta x}$. 
Because the result is analytical and because the wave 
functions of the Coulomb problem with a hard core interaction are an exponential 
multiplied by a polynomial, which is 
also the form of the 
Schr\"{o}dinger and Klein-Gordon wave functions, a complete solution of Eq.~(\ref{eq*}), 
with $V(x)\propto 1/x$ and $x>0$, can be found.   

The paper is organized as follow. In Sec.~\ref{math}, we give some useful 
mathematical results concerning the square-root operator. In Sec.~\ref{coulomb}, we solve 
the one-dimensional Coulomb problem with a hard core interaction. In Sec.~\ref{disc}, we 
compare our results to those obtained with the Schr\"{o}dinger 
equation~\cite{loud59,andr66,hain69,gome80} and with the Klein-Gordon 
equation~\cite{spec85,moss87}. At last, we give our conclusion in Sec.~\ref{conc}.

\section{Mathematical Framework}
\label{math}

In this section we give some results concerning the square-root operator which we use to 
solve the Coulomb problem supplemented by a hard core interaction.

\subsection{Integral representation of the square-root operator}
\label{theory}

To obtain the integral representation of the square-root operator we use the Fourier 
transform of the one-dimensional delta function. We have
\begin{equation}
\label{eq1}
\sqrt{-d^2_x+m^2}\ \Psi(x)=\frac{1}{2\pi}\int_{-\infty}^{+\infty}\int_{-\infty}^{+\infty} 
dp\, dq\ \sqrt{p^2+m^2}\ e^{-i(q-x)p}\ \Psi(q).
\end{equation}
Extracting the operator $-d^2_x+m^2$ and integrating over the momentum $p$ (see 
Ref.~\cite{nick84}) we obtain, 
\begin{eqnarray}
\label{eq2}
\nonumber
\sqrt{-d^2_x+m^2}\ \Psi(x)&=&\frac{1}{\pi}\left(-d^2_x+m^2\right) \int_{-\infty}^{+\infty} 
dq\ K_0(m|q-x|) \Psi(q) \\
&=&\frac{1}{\pi}\left(-d^2_x+m^2\right)\int_{0}^{+\infty} dq\ K_0(mq) [\Psi(x+q)+\Psi(x-q)],
\end{eqnarray}
where $K_0(x)$ is the modified Bessel function of order 0 \cite [p. 952]{grad80}.

\subsection{Invariant space functions of the one-dimensional square-root operator}
\label{invariant}

In this section we calculate the action of the square-root operator on the functions 
$x^n e^{-\beta x}$. We obtain that the result is equal to a polynomial of order $n$, 
$M_n(m,\beta,x)$,  
multiplied by the same exponential. Thus the space of functions $P_n(x) e^{-\beta x}$ is 
the invariant space functions of this operator. Using formula (\ref{eq2}) we have
\begin{equation}
\label{eq5}
\sqrt{-d^2_x+m^2}\ x^n e^{-\beta x}=\frac{1}{\pi}\left(-d^2_x+m^2\right) 
e^{-\beta x}\int_{0}^{+\infty} dq\ K_0(mq)\ \left[(x+q)^n e^{-\beta q}+(x-q)^n 
e^{\beta q}\right].
\end{equation}
This leads to \cite [p. 712]{grad80}
\begin{equation}
\label{eq6}
\sqrt{-d^2_x+m^2}\ x^n e^{-\beta x}=\frac{1}{\sqrt{\pi}}\left(-d^2_x+m^2\right) e^{-\beta x} 
\sum_{k=0}^{n} 
\left(\begin{array}{c}
n\\ k
\end{array}\right)
 G_k(m,\beta)\ x^{n-k}.
\end{equation}
The coefficients $G_k(m,\beta)$ are given by
\begin{equation}
\label{eq7}
G_k(m,\beta)=\frac{\Gamma(k+1)^2}{\Gamma(k+3/2)}\left(\frac{1}{(m+\beta)^{k+1}} 
F\left(k+1,1/2;k+3/2;-\frac{m-\beta}{m+\beta}\right)+(-)^k (\beta \rightarrow -\beta)\right),
\end{equation}
where $F(\alpha,\beta;\gamma;x)$ is the hypergeometric function \cite[p. 1039]{grad80}.
Performing the derivation in Eq.~(\ref{eq6}) and rearranging the obtained relation, we have:
\begin{eqnarray}
\label{eq8}
\nonumber
\sqrt{-d^2_x+m^2}\ x^n e^{-\beta x}&=&\frac{1}{\sqrt{\pi}}\ e^{-\beta x} 
\sum_{k=0}^{n}
\left(\begin{array}{c}
n\\ k
\end{array}\right)
\Bigg\{\left(m^2-\beta^2\right)G_k(m,\beta)+2\beta k G_{k-1}(m,\beta)\\ 
&-& k(k-1)G_{k-2}(m,\beta)\Bigg\}\ x^{n-k}.
\end{eqnarray}
It is possible to write the coefficients $G_k(m,\beta)$ into a more useful form. This form 
will allow us 
to find a recursion relation between the coefficients $G_k(m,\beta)$ and to simplify the 
expression (\ref{eq8}). We will be able to construct 
the polynomial, $M_n(m,\beta,x)$, for each value of $n$. We have the relation 
\cite[p. 562]{abra70}
\begin{equation}
\label{eq9}
F(a,1/2;a+1/2;-x)=\Gamma(a+1/2)\ \frac{x^{(1-2a)/4}}{\sqrt{1+x}}\ P^{1/2-a}_{-1/2}\left(
\frac{1-x}{1+x}\right),
\end{equation}
with $x>0$. The functions $P^{\mu}_{\nu}(x)$ are the associated Legendre functions 
for $x$ real and $|x|<1$. With this relation, we find that
\begin{equation}
\label{eq9bis}
G_{k-1}(m,\beta)=\frac{\Gamma(k)^2}{\sqrt{2m}\ \left(m^2-\beta^2\right)^{(2k-1)/4}}
\left[P^{1/2-k}_{-1/2}(\beta/m)\ +(-)^{k-1}\ P^{1/2-k}_{-1/2}(-\beta/m)\right].
\end{equation}
Now, using the recursion relation of the associated Legendre functions \cite[p. 1005]{grad80},
\begin{equation}
\label{eq10}
P^{\mu+2}_{\nu}(x)=-2(\mu+1)\frac{x}{\sqrt{1-x^2}}P^{\mu+1}_{\nu}(x)+(\mu-\nu)(\mu+\nu+1)
P^{\mu}_{\nu}(x),
\end{equation} 
and the explicit expression of $P^{-1/2}_{-1/2}(x)$ and $P^{1/2}_{-1/2}(x)$ 
\cite[p. 1008]{grad80}, we can write the 
following relations
\begin{equation}
\label{eq11}
G_{k+2}(m,\beta)=\frac{1}{m^2-\beta^2}\left[(k+1)^2\ G_k(m,\beta)-(2k+3)\beta\ 
G_{k+1}(m,\beta)\right],
\end{equation} 
with
\begin{eqnarray}
\label{eq12}
G_{0}(m,\beta)&=&\sqrt{\frac{\pi}{m^2-\beta^2}},\\ 
G_{1}(m,\beta)&=&-\frac{\sqrt{\pi}\ \beta}{\left(m^2-\beta^2\right)^{3/2}}.
\end{eqnarray} 
Al last, one can find, 
using Eq.~(\ref{eq11}), that the general coefficient of the sum of the Eq.~(\ref{eq8}) becomes
\begin{equation}
\label{eq13}
F_{k,n}(m,\beta)=\frac{1}{\sqrt{\pi}}\left(\begin{array}{c}
n\\ k
\end{array}\right)
\Big[\beta\ G_{k-1}(m,\beta)-(k-1)\ G_{k-2}(m,\beta)\Big] \quad \text{with}\quad k\geq 1.
\end{equation}
And with this form, a recursion relation for $F_{k,n}(m,\beta)$ can be easily found. Thus 
to conclude this section, we are able now to rewrite 
Eq.~(\ref{eq8}) into a simple form: 
\begin{eqnarray}
\label{eq14}
\sqrt{-d^2_x+m^2}\ x^n e^{-\beta x}&=& M_n(m,\beta,x)\ e^{-\beta x}=\left[\sum_{k=0}^{n} 
F_{k,n}(m,\beta)\ x^{n-k}\right] e^{-\beta x},\\
\label{eq15}
F_{0,n}(m,\beta)&=&\sqrt{m^2-\beta^2},\\
\label{eq16}
F_{1,n}(m,\beta)&=&\frac{n\beta}{\sqrt{m^2-\beta^2}},\\
\label{eq18}
\nonumber
F_{k+2,n}(m,\beta)&=&\frac{n-k-1}{(k+2)\left(m^2-\beta^2\right)}\Big[(k-1)(n-k)\ F_{k,n}
(m,\beta)\\
&-&(2k+1)\beta \ F_{k+1,n}(m,\beta)\Big], \\
\label{eq19}
F_{k,n+1}&=&\frac{n+1}{n+1-k}\ F_{k,n}.
\end{eqnarray}
We can see that we obtain the expected relation (from Eq.~(\ref{eq0})) for $n=0$. And thus 
we see that we must have $\beta < m$. With the relations (\ref{eq15})-(\ref{eq19}) the 
polynomial $M_n(m,\beta,x)$ is completely defined and we can construct it for each 
value of $n$. This result will allow us to find, with few calculations, the solution of 
the one-dimensional relativistic Coulomb problem with a hard core interaction. We give 
below the polynomials, as an example, for $n=0\rightarrow 4$. 
\begin{eqnarray}
\label{eq21bis}
M_0(m,\beta,x) &=& S, \\
M_1(m,\beta,x) &=& S\ x+\frac{\beta}{S},\\
M_2(m,\beta,x) &=& S\ x^2+\frac{2\beta}{S}\ x-\frac{m^2}{S^3},\\
M_3(m,\beta,x) &=& S\ x^3+\frac{3\beta}{S}\ x^2-\frac{3 m^2}{S^3}\ x+\frac{3 m^2 \beta}{S^5},\\
M_4(m,\beta,x) &=& S\ x^4+\frac{4\beta}{S}\ x^3-\frac{6 m^2}{S^3}\ x^2+\frac{12 m^2 \beta}{S^5}
\ x-\frac{3 m^2}{S^7} \left(m^2+4\beta^2\right),
\end{eqnarray}
with
\begin{equation}
\label{eq21ter}
S=\sqrt{m^2-\beta^2}.
\end{equation}
Note that these last relations can be simply checked by acting the square-root operator on 
each side of Eq.~(\ref{eq14}). For $n=1$, we see that we have an identity if we use the 
relation for $n=0$. Now knowing these two relations we see that the relation for $n=2$ is 
also an identity, and so on for each value of $n$.

\section{The one-dimensional relativistic Coulomb problem with a hard core interaction}
\label{coulomb}

The equation to solve is 
\begin{equation}
\label{eq22}
\sqrt{-d^2_x + m^2}\  \Psi(x) = \Bigl(E + \frac{\kappa}{x}\Bigr)\, \Psi(x).
\end{equation}
We just consider here the case $x>0$ (we will discuss after the extension to the whole $x$ 
axis). Physically this means that we have a hard core 
interaction for $x\leq 0$. Then the wave functions will possess the following asymptotic 
behavior: $\Psi(x)=0$ for $x\leq 0$ and for $x= +\infty$. Suppose that the wave functions 
have the following form
\begin{eqnarray}
\label{eq23}
\nonumber
\Psi(x) &\propto& \sum_{k=1}^{n} \gamma_{k,n}\ x^k \ e^{-\beta x} \quad \text{for} \quad x>0 
\quad \text{and} \quad n=1,2,... , \\
\Psi(x)&=&0 \quad \text{for} \quad x\leq 0.
\end{eqnarray}
We do not consider the normalization of the functions here. 
Thus replacing Eq.~(\ref{eq23}) into Eq.~(\ref{eq22}) and using Eq.~(\ref{eq14}) we obtain
\begin{equation}
\label{eq24}
\sum_{k=1}^{n}\gamma_{k,n}\sum_{p=0}^{k} F_{p,k}(m,\beta)\ x^{k-p}=E
\sum_{k=1}^{n} \gamma_{k,n}\ x^k+ \kappa \sum_{k=1}^{n} \gamma_{k,n}\ x^{k-1}.
\end{equation}
Now equaling order by order we will determine the solution. The term of order $n$ gives:
\begin{equation}
\label{eq25}
E=F_{0,n}(m,\beta)=\sqrt{m^2-\beta^2}.
\end{equation}
From the term of order $n-1$, we have:
\begin{equation}
\label{eq26}
\kappa=F_{1,n}(m,\beta),
\end{equation}
which leads to,
\begin{equation}
\label{eq26bis}
\beta=\frac{\kappa m}{n\sqrt{1+(\kappa/n)^2}}.
\end{equation}
We can remark that we have as well the necessary relation $\beta < m$.
We are now already able to determine the energy spectrum. Using 
Eq.~(\ref{eq25}) and Eq.~(\ref{eq26bis}) we have:
\begin{equation}
\label{eq27}
E=\frac{m}{\sqrt{1+(\kappa/n)^2}}.
\end{equation}

To obtain a complete solution, we must now find all the $\gamma_{k,n}$, and prove that 
the system of equations which gives these quantities has always a solution. Obviously 
we can fix $\gamma_{n,n}=1$. We see that the term of order $n-j$ determines the coefficient 
$\gamma_{n-j+1,n}$ if the previous $\gamma_{k,n}$ are known. Beginning with the term of order 
$n-2$, we obtain directly $\gamma_{n-1,n}$. And now we can get $\gamma_{n-2,n}$ from the 
term of order $n-3$. The 
independent term will fix the last factor $\gamma_{1,n}$. Thus we have a triangular system 
of $n-1$ algebraic equations with $n-1$ unknowns. This system will possess always a solution 
if the determinant of the coefficient matrix is non null. As this is a triangular matrix, the 
determinant is the product of the diagonal elements. The expression of these elements is 
$\kappa-F_{1,n-j}(m,\beta)$ which is equal to $j\beta /S$. These quantities are always non 
null since $j>0$. 
The general form of $\gamma_{k,n}$ is obtained from the term of order $n-j-1$. We have
\begin{equation}
\label{eq28bis}
\gamma_{n-j,n}= \frac{S}{j\beta} \sum_{k=0}^{j-1} \gamma_{n-k,n}\
F_{j-k+1,n-k}(m,\beta).
\end{equation} 
We can inverse the summation to finally obtain
\begin{equation}
\label{eq29}
\gamma_{n-j,n}=\sum_{p_1=0}^{j-1}\sum_{p_2=0}^{p_1-1}...\sum_{p_j=0}^{p_{j-1}-1} 
\tilde{F}(n,p_1,j)\tilde{F}(n,p_2,p_1)...\tilde{F}(n,p_j,p_{j-1}),
\end{equation} 
with
\begin{equation}
\label{eq30}
\tilde{F}(n,k,j)=\frac{S}{j\beta}\ F_{j-k+1,n-k}(m,\beta).
\end{equation} 
For the summation in Eq.~(\ref{eq29}), we must use the following rule: If in a summation over 
$p_{\alpha}$, $\alpha$ being arbitrary, the bound $p_{\alpha-1}-1$ is negative, all the 
$\tilde{F}(n,k,j)$ containing the 
indices $p_{\beta \geq \alpha}$ are equal to 1. With the formula (\ref{eq29}), we are able 
to construct the wave functions for the Coulomb problem with a hard core interaction. As 
an example we give the three first one:
\begin{equation}
\label{eq31}
\Psi(x) \propto x\ Q_n(m,\kappa,x)\ e^{-\beta x},
\end{equation}
with $\beta$ given by Eq.~(\ref{eq26bis}), and
\begin{eqnarray}
\label{eq32}
Q_1(m,\kappa,x)&=& 1,\\
Q_2(m,\kappa,x)&=& x-\frac{m^2}{S^2 \beta}, \\
Q_3(m,\kappa,x)&=& x^2-\frac{3 m^2}{S^2 \beta}\ x +
\frac{3m^2}{2\beta^2 S^4}\left(\beta^2 +m^2\right),
\end{eqnarray}
with S defined by Eq.~(\ref{eq21ter}). Again, we can perform a simple verification by 
putting these solutions into Eq.~(\ref{eq22}) and using Eq.~(\ref{eq14}).

Contrary to the Schr\"{o}dinger or Klein-Gordon equation, the extension of the solution to the 
whole $x$ axis is really more complicated. We can try to use $\exp(-\beta |x|)$ instead of 
$\exp(-\beta x)$ in our solution. But the situation is quite more difficult. Indeed, the 
construction of the solution was based on the fact that $\exp(-\beta x)$ was an eigenfunction 
of the square-root operator and that the invariant space functions of this operator was 
$P_n(x) \exp(-\beta x)$, where $P_n(x)$ is a polynomial of order $n$. But it is easy to show, 
with Eq.~(\ref{eq2}), that
\begin{equation}
\label{aaa}
\sqrt{-d_x^2+m^2}\ \exp(-m|x|)=\frac{2m}{\pi}\ K_0(m|x|).
\end{equation}
This is non null as this is the case in the Eq.~(\ref{eq14}).
Thus $\exp(-\beta |x|)$ is not an eigenfunction of the square-root operator and 
$P_n(x) \exp(-\beta |x|)$ is not an invariant space function of this operator. So it seems 
that the pure Coulomb problem has quite different solutions for the wave functions and 
certainly for the spectrum. 
\section{Discussion}
\label{disc}

The one-dimensional Coulomb problem has been treated by many authors, both 
nonrelativisticly~\cite{loud59,andr66,hain69,gome80} and 
relativisticly~\cite{spec85,moss87}. But in these works the whole $x$ axis is considered. 
As a consequence, the ground state gives some difficulties. 

In the nonrelativistic case 
the solution is:
\begin{equation}
\label{eq33}
\Psi(x)= x\ \exp(-\kappa m|x|/n)\ L_{n-1}^{(1)}(2\kappa m|x|/n),
\end{equation}
\begin{equation}
\label{eq34}
E=m-\frac{m\kappa^2}{2 n^2} \quad \text{with} \quad n=1,2,...
\end{equation}
But we see that for $n=1$, the wave function have a node at the origin. 
So this is not the wave function for the ground state. In fact it is found to be infinitely 
bounded and the wave function is a delta function~\cite{loud59,moss87}.

In the Klein-Gordon case the solution is:
\begin{equation}
\label{eq35}
\Psi(x)= x^S\ \exp(-\beta |x|/2)\ L_{n-1}^{(\gamma)}(\beta |x|),
\end{equation}
\begin{equation}
\label{eq36}
E=m/\sqrt{1+\frac{\kappa^2}{(n-1+S)^2}},
\end{equation}
with,
\begin{equation}
\label{eq37}
\beta=2m\kappa /\sqrt{(n-1+S)^2+\kappa^2} \quad \text{with} \quad n=1,2,...,
\end{equation}
and,
\begin{equation}
\label{eq38}
S=\frac{1}{2}(1+\gamma)=\frac{1}{2}\left(1\pm \sqrt{1-4\kappa^2}\right).
\end{equation}
Thus we see that we have two distinct solutions according the sign for $S$. Actually, 
spectrum with the minus sign for $S$ is not acceptable. Indeed, a reason is that, for $n=1$, 
when we perform the limit $\kappa \rightarrow 0$, we obtain $E=0$. This mean that the particle 
is still bounded when the interaction vanishes. Thus the problem for the ground state 
persists (see Ref.~\cite{moss87} for a complete discussion).

In this paper we do not consider the whole $x$ axis and we have no problem with the ground 
state. We consider a hard core interaction, for $x\leq 0$, which gives $\Psi(x)=0$ in this 
region. Thus $x \exp(-\beta x)$ is the wave function for the ground state. Actually 
the purpose of these work was to solve a particular kind of differential equations with a 
difficult to handle nonlocal operator. Indeed, any analytical solutions are known for the 
spinless Salpeter equation. Thus we do not discuss the problem of the ground state of 
the one-dimensional Coulomb problem. In Sec.~\ref{coulomb}, we have shown that the extension 
to the whole $x$ axis is not easy. Moreover, the ground state problem could persist.

To compare our result to the results of previous works, we can consider the Schr\"{o}dinger 
and the Klein-Gordon equation for the Coulomb potential supplemented by a hard core 
interaction. The spectra and the wave functions remain unchanged but the ground state problem 
has disappeared. In the three cases we have the same kind of wave functions: an 
exponential (with different arguments) multiplied by a polynomial 
(with different coefficients). 
For the spectrum we have, in the limit of small $\kappa$:
\begin{eqnarray}
\label{eq39}
E_{\text{Sch}}&=&m\, \left(1-\frac{\kappa^2}{2n^2}\right),\\
E_{\text{KG}}&=&m\, \left(1-\frac{\kappa^2}{2n^2}-\frac{\kappa^4}{n^3}+\frac{3\kappa^4}
{8n^4}\right),\\
E_{\text{Sal}}&=&m\, \left(1-\frac{\kappa^2}{2n^2}+\frac{3\kappa^4}{8n^4}\right).
\end{eqnarray}
Thus we see that in the expansion of the Salpeter spectrum the term in $n^3$ is missing compared 
to the Klein-Gordon spectrum. So the difference between these two spectra is rather 
important. For an electron in an electromagnetic Coulomb potential, the splitting is about 
$10^{-3}$ eV.

Another characteristic of the spinless Salpeter spectrum is that $\kappa$ can grow up without 
limit. This could come from the fact that we have another kinetic operator than in 
the Klein-Gordon equation and that the result could be quite different. But the main 
explanation is certainly that we do not solve the real Coulomb problem and that this 
spectrum could be different contrary to the Klein-Gordon equation, which keeps the same 
spectrum in both cases. Indeed, for the SSE, there exists a 
limit value for $\kappa$ in three dimensions~\cite{herb77}.

\section{Conclusion}
\label{conc}

The purpose of this work was to find an analytical solution of a particularly kind of 
differential equations containing a nonlocal differential operator. The equation considered 
in this paper, the 
one-dimensional spinless Salpeter equation (SSE), is a simple relativistic version of the 
one-dimensional Schr\"{o}dinger equation. The SSE is not a marginal equation. For three 
dimensions, this equation comes from the Bethe-Salpeter equation 
\cite[p. 297]{salp51,grei94}, which 
gives the correct description of bound states of two particles. Moreover,    
despite the presence of a 
so particular operator, the SSE is often used in the potential models 
(see for instance 
Refs.~\cite{stan80,cea83,gupt85,godf85,laco89,fulc94,sema97}), which give a phenomenological 
description of hadrons. 

To find this analytical solution, we calculate, in Sec. \ref{invariant}, the action of the 
square-root operator on a polynomial multiplied by an exponential and we show that this 
constitute the 
invariant space functions of this operator. To be able to perform this calculation, we have 
constructed, in Sec. \ref{theory}, an integral representation of the square-root operator. 
In Sec. \ref{coulomb}, we have obtained, without solving any differential equation, 
a complete solution of the SSE with a Coulomb potential and a hard core interaction. This 
last interaction is introduced to keep the particle in the x positive region. We remark that 
the SSE wave functions have the same form than the Schr\"{o}dinger and 
the Klein-Gordon wave functions. We remark also that the splitting between the SSE and the 
Klein-Gordon spectrum is rather important. Indeed, it is of the same order of the first 
relativistic correction given by these two equations.

\acknowledgments
We thank Dr. C. Semay, Prof. Y. Brihaye, Prof. J. Nuyts and Prof. F. Michel for 
the useful discussions.

\end{document}